\documentclass[aps,prd,showpacs,nofootinbib,preprintnumbers,twocolumn]{revtex4}
\usepackage{bm}
\usepackage{latexsym}
\usepackage{natbib}
\usepackage{url}
\usepackage{dcolumn}
\usepackage{color}
\usepackage{amsfonts,amssymb,amsmath}
\usepackage{graphicx,epsfig}
\usepackage{psfrag}
\usepackage{subfigure}
\usepackage{natbib}
\def\pbest{\texttt {Pbest}}
\def\gbest{\texttt {Gbest}}

\begin{document}
\title{Cosmological parameter estimation using Particle Swarm Optimization (PSO)}
\author{Jayanti Prasad}
\email{jayanti@iucaa.ernet.in}
\author{Tarun Souradeep}
\email{tarun@iucaa.ernet.in}
\affiliation{IUCAA, Post Bag 4, Ganeshkhind, Pune 411007, India.}
\date{\today}
\begin{abstract}
Constraining theoretical models, which are  represented by a set of parameters,
using observational data is an important exercise in cosmology. In Bayesian framework
this is done by finding  the probability distribution of parameters which best fits to the observational
data using sampling based methods like Markov Chain Monte
Carlo (MCMC).   It has been argued that MCMC may not be the best option in certain 
problems in which the target  function (likelihood)   posses local maxima or 
have very high dimensionality. Apart from this, there may be examples in which we
are mainly interested to find the point in the parameter space at which the
probability distribution has the largest value. In this situation the problem of
parameter estimation becomes an optimization problem. In the present work we
show that Particle Swarm Optimization (PSO),  which is an artificial intelligence 
inspired population based search procedure,  can also be used for cosmological
parameter estimation. Using PSO we were able to recover the best fit LCDM parameters
from the WMAP seven year data without using any prior guess value or any other
property of the probability distribution of parameters like standard deviation,
as is common in MCMC.  We also report the results of an exercise in 
which we consider a binned primordial power spectrum (to increase the
dimensionality of problem) and find that a power spectrum with features 
gives lower chi square than the standard power law. Since PSO does not 
sample the likelihood surface in a fair way, we follow a fitting procedure
to find the spread of likelihood function around the best fit point.
\end{abstract}
\pacs{}
\maketitle
\section{Introduction}
In a typical CMB data analysis pipeline first the time order data, obtained 
from an instrument like WMAP, is  reduced into a set of sky maps from which angular
power spectra are computed, and finally these spectra are reduced into a set of cosmological 
parameters representing a  model usually using Bayesian analysis
 ~\cite{1996PhRvD..54.1332J,1996PhRvL..76.1007J,2002PhRvD..66j3511L,2002PhRvD..66f3007K,2003ApJS..148..195V,2010LNP...800..147V,2003moco.book.....D,2011ApJS..192...14J,2011ApJS..192...16L,2003moco.book.....D}.
The exercise of parameter estimation involves identifying 
a set of parameters which has the highest probability of giving the observed data i.e., 
finding a point in the multidimensional parameter space at which the likelihood function
has the greatest value. Currently this exercise is done using some sampling based methods, like  
MCMC,  in which the likelihood function is sampled at discrete points, 
which are further used to compute various statistics of parameters ~\cite{2001CQGra..18.2677C,2002PhRvD..66j3511L,2010LNP...800..147V}. 
Apart from MCMC, non-sampling based methods inspired from artificial intelligence techniques, 
like artificial neural network, have also  been successfully applied in 
cosmological parameter estimation from the CMB data \cite{2007MNRAS.376L..11A}. 

In the present work we demonstrate the use of another artificial intelligence inspired
method, named Particle Swarm Optimization or PSO 
\cite{KennedEberhart1995,KennedEberhart2001,Engelbrech2002} , for cosmological parameter 
estimation using WMAP seven year data\cite{lambda}.  
Being a stochastic method, PSO also has the interesting feature that the computational cost
for searching the global maximum in the  multi-dimensional space does not grow exponentially
with the dimensionality of the search space. However, in this case also (like other stochastic methods)
the probability of convergence to the global maximum is usually guaranteed only in the asymptotic limit.
Based on a very simple idea and having very few design parameters, PSO is quite easy to program
and can provide accurate results very fast.
As compared to artificial neural network \cite{2007MNRAS.376L..11A} all the calculations in PSO 
are exact, in the sense that no extrapolation or interpolation is done at any stage i.e.,  
fitness function is computed exactly at all points.  
It has been found that PSO can outperform MCMC in certain situations,
in particular when there are a large number of local maxima present and/or the dimensionality of the 
search space is very high \cite{2010PhRvD..81f3002W}.

It has been a common practice to consider a featureless primordial power spectrum (PPS),
characterized by two parameters tilt ($n_s$) and amplitude ($A_s)$.
In this case also there is a degeneracy between the parameters of PPS
and other cosmological parameters (like $\Omega_b$),  but the likelihood surface remains 
fairly smooth and does not poses much challenge for the MCMC method.
There have been studies  which show  that a PPS with features is a better fit to
the observational data as compared to the featureless power law PPS 
\cite{2010JCAP...10..008H,2011arXiv1106.2798A,2011arXiv1109.5264M}.
 In one of such studies 
\cite{2011arXiv1109.5264M} a PPS with oscillations is considered and it is  
argued that due to additional degrees of freedom, as a result of features,
the MCMC approach is not very successful since there are a large number of local 
maxima  present in the search space and the chains frequently  tunnel from 
one local maximum to another.
In order to circumvent this problem it has been recommended first to carry out 
the search space over a subset of parameters over a grid and then use full MCMC.
PSO can be quite useful in this type of problem due to its better capabilities of
handling higher dimensional search space and large number of local maxima. 

Since the fitness function i.e., the function to be optimized, can be computed concurrently on
a parallel platform for a large number of particles, PSO  promises to give accurate
results very fast, if implemented efficiently.  

Unlike MCMC which provides the full probability 
distribution from which marginalized values of parameters
and error bars  can be  computed, PSO just gives the location of the best fit point, called the \gbest, 
(definition in next section) and one needs to find some way to compute error bars. 
In the present work we fit the effective 
chi square ($\chi^2_{\rm{eff}} = -2 \log {\mathcal L}$) by a multi-dimensional paraboloid 
around the best fit point  and compute the error bars from the fitting coefficients of that.
 On the basis of
the errors we set the range for the two dimensional grids which we consider for
various combinations of parameters (keeping all other parameters fixed
to their best fit values)  and make contour plots which show not only the extent
to which the likelihood surface is spread (errors) but also  the correlation 
between various parameters. These contour plots also confirm that PSO 
does find the global maximum of the likelihood function.

We observe that in most cases  (which we have considered) not only does the 
\gbest~ approach towards the best fit point, the average location of particles also
approaches towards it, as PSO progresses. The average location of PSO particles in the 
multi-dimensional search space can be used to check the robustness of PSO.
Since PSO is designed mainly for finding the best fit point, therefore the sampling 
done by PSO particles is not designed to be a fair  representation of the likelihood surface.

The plan of the paper is as follows. In~\textsection~\ref{ss:pso_formalism}
we discuss a simple implementation of the Particle Swarm Optimization in detail.
In particular, we focus on the dynamics of particles, setting-up initial conditions, 
boundary conditions and the convergence criteria. We also define all the design parameters and
variables of PSO in~\textsection~\ref{ss:pso_formalism}. 
A very brief overview of parameter estimation in Bayesian formalism is discussed in 
~\textsection \ref{ss:param_est}. In place of discussing the full prescription of
Maximum Likelihood (ML) estimation, we mainly focus on the computation of error bars from covariance 
or Hessian matrix in this section. We present our results of parameter estimation using PSO for 
WMAP seven year data in~\textsection~\ref{ss:pso_results}. Apart from giving the 
best fit parameters and errors which we get from our fitting exercise, we also
present contour plots for various combinations of the parameters.
Along with the best fit parameter estimates which we get from PSO, we give a comparison 
of the results obtained from PSO and as are reported by WMAP team using MCMC. 
In~\textsection \ref{ss:pso_summary} we summarize the PSO and discuss  its advantages 
and disadvantages.
\section{Particle Swarm Optimization}
\label{ss:pso_formalism}
Formally proposed by James Kennedy and Russell Eberhart in 1995 
\cite{KennedEberhart1995}, PSO has been
successfully tested and applied in many engineering and artificial intelligence
problems \cite{Lazinic2009,Boeringer2004,Robinson2004}.
Recently it has been applied in astrophysical problems also 
\cite{2005MNRAS.359..251S,2010PhRvD..81f3002W,2011ApJ...727...80R}.
In PSO, a set of ``particles"  driven by  ``cognition" and ``social" factors explore the 
multi-dimensional search space by carrying out random walks, determined by a set of ``design parameters".
The accuracy and performance of the algorithm depends on the values of the design parameters
as well as the function to be optimized i.e., the fitness function. 

Particle Swarm Optimization is based on observations of social dynamics, bird flocks,
fish schools and other forms of group behavior. Personal discoveries made by
members of the group and shared with everyone else can help everyone to become more
efficient in making new discoveries.
Efficient personal search and communication with other members of the group
can lead to rapid success for the group in search of some common goal i.e., food etc.

At present there exists  many implementations of the Particle Swarm Optimization.
Here we consider one of the simplest forms \cite{KennedEberhart1995} the elements of which 
are shared by many other implementations. 
Before describing the working of our PSO implementation, it is useful to define 
some of the key terms which are used to describe PSO.
\begin{enumerate}
\item{{\bf\ Particles:-}} The term ``particles" in PSO is used for  ``computational agents" and has 
no relation with any form of physical particles.
PSO particles  have no mass and occupy no volume (however, they can have  
weights called inertia weights which will be discussed later).
PSO particles are distinguished from each other on the basis of their identification numbers or 
{\it\ id}s  and have ``positions" and ``velocities".
In our discussion we represent the position and velocity of a particle with id {\it\  i}
at step (``time") $t$  by vector $X^i(t)$ and $V^i(t)$ respectively.
{\item{\bf\ Fitness function:-}} The function ${\mathcal F}(X)$ to be used for searching the global maximum 
is called the ``fitness function" or ``optimization function". In the present case we 
use $-2 \log {\mathcal L}$ or $\chi^2_{\rm{eff}}$ as our optimization function, where ${\mathcal L}$ is the 
CMB likelihood function. 
{\item {\bf\ \pbest :-}} The maximum value of the optimization function ${\mathcal F}^i(t)$  for a particle $i$ 
till the present step  (time $N_t$)  is called  its $\text{\pbest}^i$.
\begin{equation}
\text{\pbest}^i=\text{Max}\{{\mathcal F}^i(t),t=0,1,2...,N_t\}
\end{equation}
the location of the $\text{\pbest}^i$~is represented by the vector $P^i$
\begin{equation}
 P^i=X^{i}(t)~~\text{if}~~{\mathcal F}^i(t)= \text{\pbest}^i 
\end{equation}
{\item{\bf\ \gbest :-}} The largest value of \pbest~  among all particles is called the ~~\gbest. The value of
\gbest~  changes only when any of the particles finds a new position at which the value of the 
fitness function is larger than any of the earlier values.
\begin{equation}
\text{\gbest} =\text{Max}\{\text{\pbest}^i,i=0,1,2...,N_{p} \}.
\end{equation}
Here $N_{p}$ is the number of particles. The location of the ~\gbest~ is given by the vector $G$.
\begin{equation}
G=X^i(t)~~\text{if}~~\text{\pbest}^i=\text{\gbest}.
\end{equation}
\end{enumerate}
\subsection{Dynamics of PSO particles}
The following equation is used to update the positions of the particles~\cite{KennedEberhart1995},
\begin{equation}
X^i(t+1) = X^i(t) + V^i(t+1),
\label{xevolve}
\end{equation}
where velocity $V^i(t+1)$ for the particle $i$ at step  $(t+1)$ is computed in the
following way
\begin{eqnarray}
V^i(t+1) = w V^i(t) + c_1 \xi_1 \left \{X^i(t)-P^i \right \}  \nonumber \\ 
+  c_2 \xi_2  \left \{ X^i(t)-G\right \}.
\label{veq}
\end{eqnarray}
Here  $c_1$ and $c_2$ are called acceleration coefficients, $w$ is called the inertia weight
and $\xi_1$ and $\xi_2$ are two uniform random numbers in the range $[0,1]$.
The values of the acceleration coefficients $c_1$ and  $c_2$ decide the contribution due to personal (cognitive)
learning and social learning respectively.

The first factor in the right hand side of equation~ (\ref{veq}) moves the particle along a straight line
and the second and third factors accelerate it towards the location of ~\pbest~ and  ~\gbest~ respectively.
Kennedy  and Eberhart \cite{KennedEberhart1995} use $c_1=c_2=2$ to give it a mean of unity, so that the particle
would overfly  the target about half of the time. 

Although the equation~(\ref{veq}) is most commonly used, the following 
version is also in use \cite{2005MNRAS.359..251S}.
\begin{align}
V^i(t+1)= K \left \{V^i(t)+  C_1 \xi_1 \left(X^i(t)-P^i\right) \right. \nonumber \\ 
\left.+ C_2 \xi_2 \left(X^i(t)- G \right)\right \}
\label{veq2}
\end{align}
where $K$ is called the constriction factor and is defined in the following way:
\begin{equation}
K = \frac{2}{|2-\phi - \sqrt{\phi^2-4\phi}|}
\end{equation}
where $\phi=C_1+C_2$. Here the recommended values are $C_1=C_2=2.05$ which gives $K=0.729$.  Equation
(\ref{veq2}) is equivalent to equation ~(\ref{veq}) with $c_1=KC_1,c_2=KC_2$ and $w=K$. Since there are
many implementations (with different values of design parameters or with some
new parameters) we have decided to work with PSO standard 2006 \cite{pso2006}  which uses the following
values  of the design parameters in equation (\ref{veq})
\begin{equation}
w = \frac{1}{2 \log (2)} = 0.72 
\end{equation}
and 
\begin{equation}
c_1 = c_2 = 0.5 + \log (2) = 1.193
\end{equation}
since we were able to get quite accurate results with the  values of design parameters suggested in
PSO standard 2006\cite{pso2006},  we decided to adopt these values in our implementation. 
We did try a few other values but could not find any significant improvement.

In particle swarm optimization all the particles can communicate with each other or the communication
can be restricted between only subsets of particles. The first case is found to be more useful for
intensive local search and the second one for global search. In our implementation we let every
particle share the information about its ~\pbest~ with every other particle.

\subsection{Maximum Velocity}
In order to stop particles leaving the search space we need to limit the maximum 
velocity which particles can acquire. This can be done by setting the maximum  
velocity along various dimensions. It has been a common practice to keep 
the  maximum velocity proportional to the search range. 
\begin{equation}
V^i(t) =
\begin{cases}
  V_{max}, &\text{if}~~ V^i(t) >  V_{max} \\
 -V_{max}, &\text{if}~~ V^i(t) < -V_{max} 
\end{cases}
\end{equation}
where $V_{max}$ is also a design parameter. We use $V_{max}=c_v (X_{max}-X_{min})$ with
$c_v=0.5$ where $[X_{min},X_{max}]$ is  our search range. This means that the biggest jump
a particle can make is half of the size of the search range.
\subsection{Initial Conditions}
We assign random positions and velocities to
particles in the beginning.
\begin{equation}
X^i(t=0) = X_{min} + \xi \times (X_{max}-X_{min})   
\end{equation}
and
\begin{equation}
V^i(t=0) = \xi V_{max},    
\end{equation}
where $\xi$ is a uniform random number in the range $[0-1]$. 
Apart from the above initial conditions ``particles on a grid" initial condition can also be  used.
From our trial runs we have found that the final outcome i.e., the location of the \gbest~ is
not very sensitive to the initial condition. 
\subsection{Boundary condition}
We use the ``reflecting wall" boundary condition in which a particle
reverses its velocity component perpendicular to the boundary  when  it tries  to cross  the boundary.
\begin{equation}
V^i(t) = -V^i(t) 
\end{equation}
and 
\begin{equation}
\begin{cases}
X^i(t)=X_{max} &\text{when}~~ X^i(t)  >  X_{max} \\
X^i(t)=X_{min} &\text{when}~~ X^i(t)  <  X_{min} 
\end{cases}
\end{equation}
\subsection{Termination criteria}
PSO particle trajectories are like ``chains" in MCMC, however, they are coupled 
to each other by  second acceleration coefficient $c_2$. In the limit of $c_2=0$ particle do 
not exchange information but in that case PSO will become meaningless. We have used Gelman-Rubin 
R statistics\cite{1992GelmanRubin,1998BrooksGelman} in order to find out when the exploration by PSO particles should be stopped.
In order to use Gelman-Rubin statistics we use the mean of variance $W$ within PSO
particle trajectories and variance of the mean $B$ across PSO trajectories.

At any time $N_t$ the mean value the trajectory of a PSO particle is
\begin{equation}
{\bar X}^i = \sum\limits_{j=1}^{N_t} X^i(t)
\end{equation}
and the variance 
\begin{equation}
\sigma^2_i = \frac{1}{N_t-1} \sum\limits_{j=1}^{N_t}(X^i(t)-{\bar X}^i)^2 
\end{equation}
and the mean of variance 
\begin{equation}
W=\frac{1}{N_p} \sum\limits_{i=1}^{N_p} \sigma_i^2
\end{equation}
In order to compute variance of means we firstly compute mean of means
\begin{equation}
Y  = \sum\limits_{i=1}^{N_p} {\bar X}^i 
\end{equation}
and then 
\begin{equation}
B = \frac{N_t}{N_p-1}\sum\limits_{i=1}^{N_p}({\bar X}_i - Y)^2 
\end{equation}
The variance of stationary distribution can be written as weighted sum of 
$W$ and $B$.
\begin{equation}
Z=\left(1-\frac{1}{N_t}  \right) W + \frac{1}{N_t}B 
\end{equation}

The potential scale reduction factor ${\hat R}$ is given by 
\begin{equation}
{\hat R} = \sqrt{\frac{Z}{W}}
\end{equation}
In general when the value of ${\hat R}$ as low as 1 we can assume that the 
convergence has been obtained. 
\begin{table}
\begin{center}
\begin{tabular}{|c|l|c|} \hline \hline 
Parameter & Description & Value  \\ \hline \hline 
$w$       & Inertia weight &  0.72 \\ \hline
$c_1$     & Acceleration parameters (personal) & 1.193 \\ \hline
$c_1$     & Acceleration parameters (social)   & 1.193 \\ \hline
$c_v=V_{max}/\Delta X$ & maximum velocity parameter & 0.5 \\ \hline
$N_{p}$ & Number of particles & 30 \\ \hline
$N_{d}$ & Search dimensions  & 6  \\ \hline
\end{tabular}
\caption{PSO design parameters}
\label{psopara}
\end{center}
\end{table}
\section{Cosmological Parameter estimation}
\label{ss:param_est}
Cosmic Microwave Background  temperature and polarization anisotropies observed in 
the sky represent the fluctuations in the baryon-photon fluid 
at the epoch of last scattering i.e., when  electrons were last scattered by photons  before they combined with protons and 
formed hydrogen atoms,  contain a lot of information about the cosmological parameters 
\cite{1984ApJ...285L..45B,2004astro.ph..3344C,1996LNP...470..207H,2003moco.book.....D}.
Due to the Gaussian nature of the density fluctuations at the epoch of last scattering (primordial
fluctuations) as predicted by inflationary models, the most important information about
the cosmological parameters is encoded in the angular two point correlation function or power 
spectrum. 

It is a common practice to represent the temperature anisotropies in the CMB sky in 
spherical harmonic expansion 
\begin{equation}
\frac{\Delta T ({\hat n})}{T_0} = \sum\limits_{lm}  a_{lm} Y_{lm}({\hat n}) 
\label{tmap}
\end{equation}
where $T_0$ is the average temperature i.e., the monopole term. The angular two point 
correlation function is given by
\begin{equation}
C(\theta) = \left < \frac{\Delta T ({\hat n})}{T_0} \frac{\Delta T ({\hat n}')}{T_0} \right > =  
\sum\limits_{l} \frac{2l+1}{4\pi} C_l P_l{\cos\theta}
\end{equation}
where $\theta$ is angle between directions ${\hat n}$  and ${\hat n'}$ in the sky. 
The angular power spectrum $C_l$ is defined as 
\begin{equation}
C_l = \left < a_{lm} a^{*}_{lm} \right > =  \left < |a_{lm}|^2 \right >.  
\label{cl}
\end{equation} 

As mentioned above, the angular power spectrum $C_l$ (or angular two point correlation function $C(\theta)$)
depends on a large  number of cosmological  parameters representing various energy densities in the universe, 
primordial fluctuations, and the  physical processes relevant in the early universe like reionization and 
recombination \cite{1984ApJ...285L..45B,1987MNRAS.226..655B,1997MNRAS.291L..33B,1998PhRvD..57.2117B}. 
Many of the cosmological parameters affect the angular power spectrum in the same way, i.e.,  
have degeneracies. However, it is possible to form a set of parameters, called ``normal parameters" which
affect the angular power spectrum in a unique way \cite{2003ApJ...596..725C,1997MNRAS.291L..33B}.
The most common cosmological parameters which have been used 
to fit observational $C_l$ (we also use these) are density parameters for cold dark matter ($\Omega_c h^2$), 
baryons $(\Omega_b H^2)$, cosmological constant $(\Omega_{\Lambda})$, primordial scalar power 
spectrum index $(n_s)$, and normalization $(A_s)$, and reionization optical depth $(\tau)$.
In our analysis we do not consider the ``Hubble parameter" $h$ as a free parameter and compute 
its  value from other parameters for a spatially flat model.
\begin{equation}
h=\sqrt{\frac{\Omega_bh^2+\Omega_ch^2}{1-\Omega_{\Lambda}}}
\label{eqn:hval}
\end{equation}

It is now a common practice to follow a ``~line~of~sight~" integration approach for computing the 
angular power spectrum $C_l$ for a set of cosmological parameters which is a computationally expensive process.
The publicly available code CAMB ~\cite{camb,2000ApJ...538..473L} is based on an earlier code named 
CMBFAST ~\cite{1996ApJ...469..437S} which can compute the angular power spectrum $C_l$ on a shared 
memory platform in a short time. 
\subsection{Bayesian Analysis}
In the framework of Bayesian analysis the probability of obtaining a set of parameters $\Theta$
which is consistent with a data set $D$ for a given prior $I$ is given by 
\begin{equation}
P(\Theta |D,I) = \frac{P(D|\Theta,I) P(\Theta | I)}{P(D|I)}.
\end{equation}
In the above equation $P(\Theta |D,I)$ is the posterior probability distribution, 
$ P(D|\Theta,I)$ is the likelihood function (which will be represented by ${\mathcal L}$) and
$ P(\Theta | I)$ is the prior. The denominator $P(D|I)$ called ``evidence" 
is used for the purpose of normalization and does not depend on the parameters $\Theta$ so
it can be ignored for the present purpose.

The likelihood function for  a CMB experiment with $N_p$ pixels  is given by  
\cite{2003moco.book.....D}
\begin{equation}
{\mathcal L}(\Delta |\Theta) =  \frac{1}{(2\pi)^{N_p/2}} \frac{1}{|C|^{N_p/2}}  
\exp \left [-\frac{1}{2} \Delta  C^{-1} \Delta \right ]
\end{equation}
where ~$\Delta$~ represents an estimator of the observed data vector, having $N_p$ entries, and 
$C$ is the joint covariance matrix i.e., sum of the signal and noise covariance matrix 
\begin{equation}
C=S+N.
\end{equation}

The noise covariance matrix $N$ can be approximated by a diagonal matrix and the signal covariance matrix 
$S$ is given by \cite{2010LNP...800..147V}  
\begin{equation}
S_{ij} =\sum\limits_{l}\frac{2l+1}{4\pi} C_l P_l(\cos \theta) 
\end{equation}
where $\theta$ is the angle between the directions ${\hat n}_i$ and ${\hat n}_j$
representing pixel $i$ and pixel $j$ respectively.

Exact likelihood computation by a brute force method is computationally expensive since it
involves inversion of a $N_p\times N_p$ matrix which is a formidable task for an experiment 
with very large number of pixels. Many approximations   have been proposed 
which reduce the cost of likelihood computation 
significantly~\cite{2008PhRvD..77j3013H,2009PhRvD..79h3012H,2010LNP...800..147V}. 

In the present work we use the likelihood code provided by the WMAP team for
computing the likelihood function, which takes the theoretical angular power spectrum computed 
by CAMB,  and the power spectrum estimated by the WMAP experiment\cite{wmap}, as inputs. 

The exercise of obtaining the best fit cosmological parameters involves finding a 
point in the multidimensional parameter space,  at which the value of the likelihood
function ${\mathcal L}$ is maximum or  $-2\log {\mathcal L}$ is minimum. 
Apart from the best fit values one is also interested in the error bars  on 
the estimated parameters which involves knowing the shape of the likelihood 
function around the best fit values, for which we follow a fitting procedure as discussed below. 

\subsection{Likelihood fitting}
\label{ss:lklft}
Close to the best fit point we can approximate the likelihood function Gaussian :
\begin{equation}
{\mathcal L} = {\mathcal L}_0 \exp[-\frac{1}{2} \Delta^T R \Delta]
\end{equation}
where $\Delta_i = \theta_i - \theta_{i,0}$ where $\theta_{i,0}$ is the maximum likelihood
value of the parameter $\theta_i$ and $R$ is the curvature matrix. The covariance matrix
$C$ which is the  inverse of curvature matrix $R$ gives an estimate of the standard errors
which maximum likelihood fitting can give \cite{1996PhRvD..54.1332J}. 

The standard error $\Delta \theta_i $ in parameter $\theta_i$ is given by:
\begin{equation}
\Delta \theta_i = \sqrt{C_{ii}} = 1/\sqrt{ [(R)^{-1}]_{ii}}
\end{equation}

We fit $-2 (\log {\mathcal L} - \log {\mathcal L}_0)=\Delta \chi^2_{\rm{eff}}$
with a paraboloid and compute the coefficients of fitting and identify those with the 
elements of curvature matrix. 
\section{Results}
\label{ss:pso_results}
We compute the best fit cosmological parameters from the WMAP seven year data for
a six parameter  model with model parameters $\Omega_b h^2, \Omega_c h^2, 
\Omega_{\Lambda}, n_s, A_s$ and $\tau$ using PSO.
Before presenting our results quantitatively, we consider it useful to present a 
qualitative comparison of the way parameters are estimated in the Markov-Chain Monte 
Carlo methods and in Particle Swarm Optimization.
In particular, we want to highlight the way parameter space is explored and sampled
in  PSO and MCMC methods.
\subsection{Markov Chain and PSO exploration}
\begin{figure}[t]
\begin{center}
\epsfig{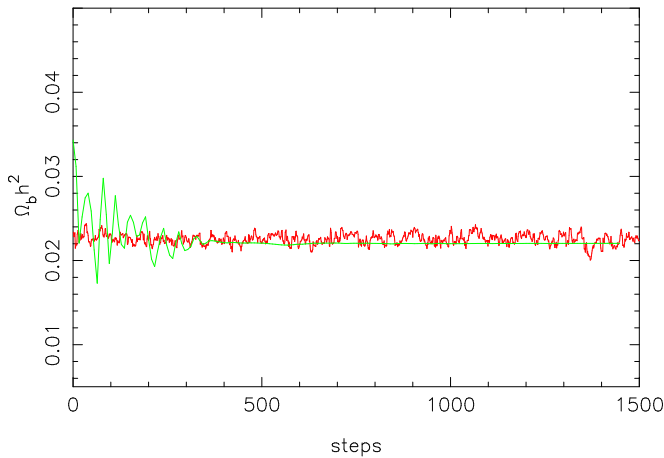}
\caption{In this figure the red line shows a Markov chain which has been obtained 
from a typical run of COSMOMC and the green line shows the trajectory of a PSO
particle,  along the same dimension i.e. $\Omega_b h^2$.
Markov chain as well as a PSO trajectory can begin anywhere in the range 
and progressively move towards the best fit location.
However, in the case of PSO  the particle approaches towards the best fit location (\gbest)
in an oscillatory manor with successively decreasing amplitude, which is not the case 
for a Markov Chain since its  step size does not vary much. 
Only after a sufficient number of PSO steps the particle positions and the Markov 
chain converge. Since there are more number of points for the Markov chain as 
compared to the PSO,  we use x-scale such that we have five Markov points for every  PSO
point.}
\label{fig:chains}
\end{center}
\end{figure}
The nature of exploration by a Markov chain and that  by a set of PSO particles is
completely different. 
However, there are some similarities also, for example, 
in both the cases the random walk is governed by the optimization function or the 
fitness function. In MCMC, the proposal density is directly related to the function to be sampled. 
In the case of MCMC exploration of a chain is completely local,  in the 
sense that whether a step will be selected or rejected depends only on 
the values of the fitness function at the current location and the next location.
However, in the case of PSO, particles always have some information 
about the global maximum ~\gbest~, which keeps changing.
In general the step size  does not change in MCMC, however, in the case of PSO it
rapidly falls as ~\gbest~ approaches close to the global maximum.

\begin{figure}
\begin{center}
\epsfig{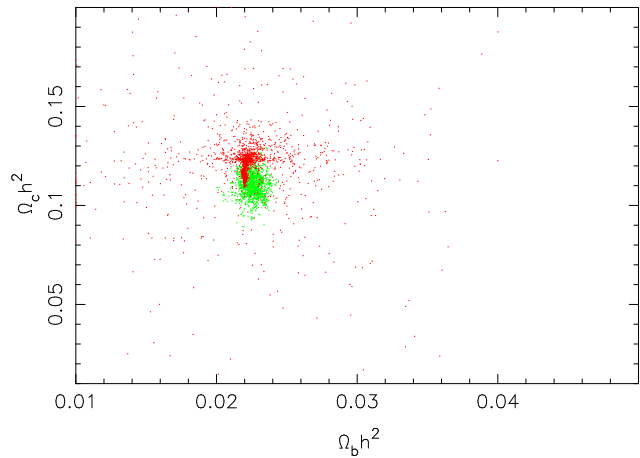} 
\caption{In this figure the red and the green points show the distribution of the positions
of PSO particles and samples from a Markov chain respectively, in the same plane.
From the figure it can be noticed that in the initial stage  the scatter of PSO particles is very large
(see Figure~\ref{fig:chains} also), however, close to the convergence
all particles get confined in a very compact region.
The distribution of the sample points in the case of Markov chain is much more symmetric than in PSO.
We suspect that this is due to the different role played by the stochastic variables
(random numbers) in PSO as compared to that in the Markov chains. The non-symmetric distribution makes 
PSO less favorable if we want to find the shape of the likelihood close to the best fit values 
(in order to report errors) in comparison to Markov Chain.}
\label{fig:psomc2d}
\end{center}
\end{figure}

Markov chains sample the fitness function in the multi-dimensional parameter 
space using methods like Metropolis Hastings. 
The sampling algorithm ensures that more points are sampled from regions in 
which the fitness function has large values and less points from the regions
in which it has small values. 
The values of best fit parameters are obtained after marginalization.  
In case of PSO a set of ``particles" explore the multidimensional space guided by their personal  
i.e., \pbest~ and collective i.e.,~\gbest~ discoveries (see equation~(\ref{xevolve}) 
and equation~(\ref{veq})). The progress of a chain in MCMC and trajectory of 
a particle in PSO are  very different. 
Starting from any arbitrary point in the multi-dimensional
space both approach towards the region where the probability of global maximum is
high, however, the way they approach is  different. 

Figure~\ref{fig:chains} shows a typical Markov chain and the trajectory of a PSO
particle. Since there were greater number of steps in the Markov chain than in PSO, we 
have stretched the x-axis for PSO trajectory by a factor of five, i.e, there are five 
Markov chain points for every PSO point for the same range on the x-axis. 
From the figure it can be noticed  that the PSO particle reaches the global maximum
by performing oscillatory motion with gradually decreasing amplitude. However, in the
case of Markov chain the progress is very smooth.   

One of the most common ways to present the results of a parameter estimation exercise is to 
make two dimensional scatter or contour plots. In MCMC it is done by marginalizing the 
sampled function along all other dimensions apart from the two for which we want scatter 
or contour plots. For a general case, the location of the point at which the likelihood
function peaks may be different from the average location computed on the basis of
the one dimensional probability distribution obtained after marginalization.
In the case of PSO also, we present the average location  of the particles,
apart from finding the point at which the likelihood function peaks. 

The red and green points in  Figure~\ref{fig:psomc2d} show the projection of the positions 
of PSO particles and a set of sample points from a Markov chain in a 
two-dimensional plane of the parameter space, respectively.
From the figure it can be noticed that although the sampled points in both cases 
cluster around the same point (\gbest~ in PSO) the distribution are completely different. 
In particular the points are more symmetrically distributed around the best fit value 
in the Markov chain case  as compared to that in  PSO.
Since PSO results are always quoted in term of ~\gbest~ therefore the distribution of points 
does not change the results in any way. However, since in the present work we make explicit
use of the PSO particle distribution also (for  fitting the likelihood function for computing errors),
it  can create some problem.
\subsection{Best fit cosmological parameters}
\begin{figure}
\begin{center}
\epsfig{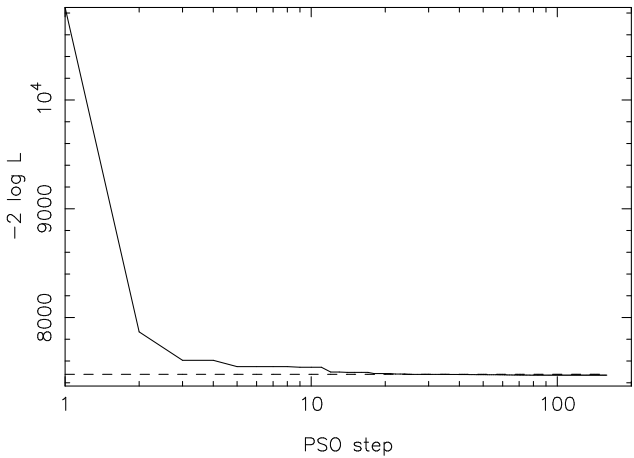}
\caption{The solid line in the above figure  shows the change in the fitness function $-2 \log {\mathcal L}$ 
as PSO steps progress, and the dashed line shows the value for  the WMAP seven year data. 
From this figure it can be noticed that in the beginning improvement in the value of the 
fitness function is quite rapid, but after some time it saturates, primarily  because 
once the particles reach close to the global maximum as given by the \gbest~ their velocities drop.}
\label{fig:gbest}
\end{center}
\end{figure}
In order to test the working of our PSO module we considered a six dimensional cosmological
model $(\Omega_bh^2,\Omega_ch^2,\Omega_{\Lambda},n_s,A_s,\tau)$ (see Figures~1 of \cite{2011ApJS..192...16L})
and tried to estimate its parameters from the WMAP seven year data. 
The range over which we tried to optimize our fitness function and the results are
given in Table~(\ref{tab:results}).
\begin{table*}
\begin{center}
\begin{tabular}{|l|l||l|l||l|l|l|l|}
\hline \hline
\multicolumn{8}{|c|}{\bf\ Cosmological parameters from PSO} \\ \hline
\multicolumn{2}{|c|}{Variable} & \multicolumn{3}{c|}{PSO best fit}&\multicolumn{2}{l|}{WMAP best fit\cite{2011ApJS..192...16L}}
 & Difference (\gbest -ML) \\ \hline
\cline{2-5}   &  Range &         \gbest ($\chi^2_{\rm{eff}}=7469.73$) & Mean         & std dev.    &  ML($\chi^2_{\rm{eff}}=7486.57$)   & Mean  &     \\ \hline\hline
$\Omega_b h^2$ & (0.01,0.04)    & 0.022036   &  0.022030   & 0.000456   &  0.02227  &  $0.02249^{0.00056}_{-0.00057}$ & -0.000234(-1.05\%) \\ \hline
$\Omega_c h^2$ & (0.01,0.20)    &  0.112313     & 0.112435   &  0.005276   & 0.1116  &    $0.1120\pm 0.0056$  &   0.000713 (0.63\%)  \\ \hline
$\Omega_{\Lambda} $ &(0.50,0.75) & 0.721896   & 0.720353   &  0.029047   & 0.729    &   $0.727 ^{+0.030}_{-0.029}$  & -0.007104(-0.97\%) \\ \hline
$n_s $ & (0.50,1.50)            & 0.963512     & 0.963278  & 0.011730    & 0.966   & $0.967\pm 0.014 $  &    -0.002488(-0.25\%) \\ \hline
$A_s/10^{-9} $ & (1.0,4.0)      &2.448498    &2.454202   &   0.106615   & 2.42      & $2.43\pm 0.11$ &        0.028498(1.17\%)\\ \hline
$\tau  $ & (0.01,0.11)          &0.08009 & 0.083930  &   0.012113   &0.0865    & $0.088 \pm 0.015 $ &       -0.00641(-7.41\%) \\ \hline \hline
\end{tabular}
\caption{The first column in the above table shows the PSO fitting parameters 
and the second, third, fourth  and fifth columns show the search range, the location of  ~\gbest, the 
average position of PSO particles and the error (which is computed by fitting the sampled function) respectively. 
In the sixth and seventh columns we give the best fit (ML) and the average values of the cosmological 
parameters derived from WMAP seven years likelihood estimation respectively. 
In the last column we give the difference between our best fit parameters (PSO parameters) and WMAP team's  
best fit parameters (difference between ML and \gbest~ values).
From this table it is clear that roughly there is good agreement between the PSO best fit parameters and WMAP 
team's best fit parameters from the seven year data.}
\label{tab:results}
\end{center}
\end{table*}

In  Figure~\ref{fig:gbest} we show the evolution of the fitness function ($-2 \log {\mathcal L}$)
with PSO steps.  We also show  value for the fitness function for WMAP seven years best fit cosmological
parameters (dashed line). From the figure it can be noticed that the value obtained by PSO finally 
converges to the WMAP seven year value.
Since the velocity with which particles move toward ~\gbest~ is proportional to their distance from  ~\gbest,
we get large jumps in the fitness function in the beginning.
\begin{figure*}
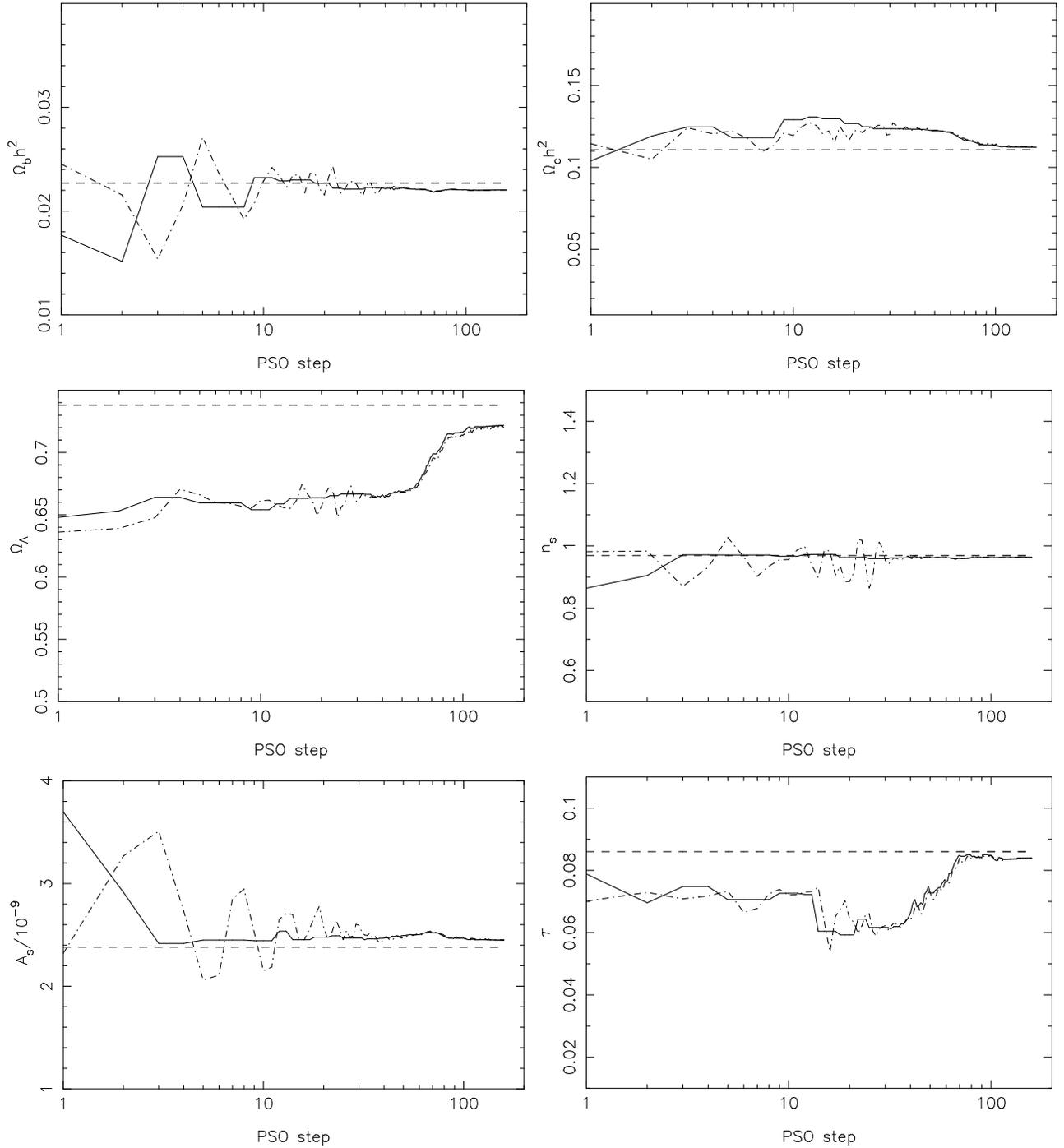

\begin{center}
\begin{tabular}{cc}
\epsfig{width=6cm,angle=-90,file=gbest_1.ps} &
\epsfig{width=6cm,angle=-90,file=gbest_2.ps} \\
\epsfig{width=6cm,angle=-90,file=gbest_3.ps} &
\epsfig{width=6cm,angle=-90,file=gbest_4.ps} \\
\epsfig{width=6cm,angle=-90,file=gbest_5.ps} &
\epsfig{width=6cm,angle=-90,file=gbest_6.ps}
\end{tabular}
\caption{In each  panel of the above figure we show the location of the best fit points
(location of \gbest) and the average location of PSO particles, by the solid and dot-dashed 
lines respectively. We also show the best fit values given by the WMAP team by dashed lines.
From the above figure it can be noticed that as PSO progresses the average location of
PSO particles and the location of the \gbest~ converge, which can be used as a robust check.
For most cases the best fit values obtained by PSO match well with standard LCDM values, 
but there are some differences also (see table \ref{tab:results}). }
\label{fig:gbest_para}
\end{center}
\end{figure*}

In PSO the values of the best fit parameters, location in the parameter space at which 
the likelihood function peaks, is represented by ~\gbest. 
For consistency and robustness we not only give the location of the ~\gbest, 
we also give the average location of the PSO  particles. 
It is not a surprise that as PSO progresses,  the average position of PSO particles and
the location of the best fit point converge to \gbest. 
In a case when there are local maxima also present, some of the PSO particles may trap in these,
but, the  average location of particles still follows  ~\gbest.
In Figure~\ref{fig:gbest_para} we show the location of  \gbest~ and the average position 
of the PSO particles in our six dimensional search space at  different steps.
Note that in our model $h$ is not a fitting parameter, we get 
its value from the flatness condition (see equation~(\ref{eqn:hval})).

\begin{figure}
\begin{center}
\epsfig{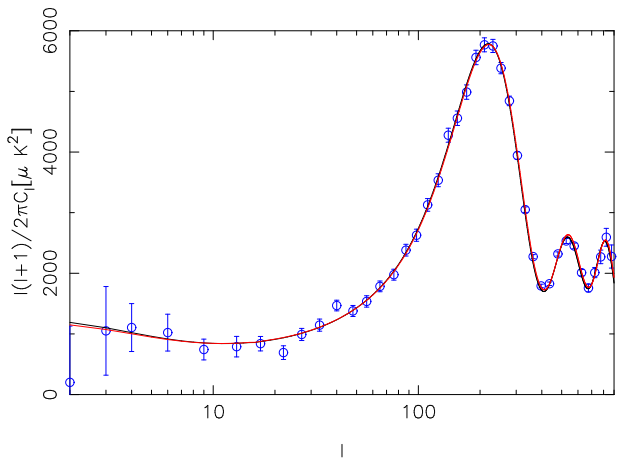}
\caption{The red, black and blue lines in the above figure represent the best fit angular power spectrum
recovered from PSO, standard LCDM power spectrum  and the binned power spectrum of WMAP seven year data respectively.  Note that the
PSO best fit angular power spectrum is very close to that provided by the WMAP team. The small difference
(see table~(\ref{tab:results}))  between the PSO best fit parameters and the WMAP best fit parameters leads
to a difference of 7 in $\chi^2_{\rm{eff}}$ (it is smaller by 7 for PSO).}
\label{fig:cls}
\end{center}
\end{figure}

The black, red and blue lines in Figure~\ref{fig:cls} show the best fit 
angular power spectrum obtained by MCMC analysis, from PSO code and the 
binned power spectrum (with error bars)  provided by the WMAP team for
the seven year data. From the figure it is clear that the power spectrum 
which we obtain from our PSO code closely follows other two curves. 

\subsection{Error estimates}
In order to compute error bars on the parameters estimated using PSO, 
we fit (as discussed in \textsection~\ref{ss:lklft}) a
six dimensional paraboloid to a subset of sampled points to $\Delta \chi^2_{\rm{eff}}
= - 2 \log {\mathcal L} - (-2 \log {\mathcal L}_0)$  where ${\mathcal L_0}$ is the value of 
the likelihood function 
or \gbest at the last PSO step.

\begin{equation}
\Delta \chi^2_{\rm{eff}} = [{\tilde \Theta}] [\alpha] [{\tilde \Theta}]^T
\end{equation}
where $\Theta=(\Omega_b h^2,\Omega_c h^2, \Omega_{\Lambda}, n_s, A_s, \tau)$ 
and $[\alpha]$ is a  $6\times 6$ symmetric matrix with 21 independent coefficients. 

The six dimensional vector ${\tilde \Theta}$ is defined as
\begin{equation}
{\tilde \Theta } = \frac{\Theta - \Theta_{\gbest}} {\Theta_{\gbest}}.
\end{equation}

We used multi-parameter fitting subroutine of GNU scientific library for the fitting \cite{gsl}.
In order to limit our fitting to a subset of points, we consider only those points
which are within a six dimensional hyper-sphere of radius i.e., $|{\tilde \theta_i}| < 0.1$
where $\theta_i$ is a component of the vector ${\tilde \Theta}$, and $\Delta \chi^2_{\rm{eff}} < 10$.

After obtaining the fitting  matrix $[\alpha]$ we invert it and compute the covariance matrix 
$[C]=[\alpha]^{-1}$ and compute error for the parameter $\theta_i$ from it 
\begin{equation}
\Delta \theta_i = \sqrt{C_{ii}} \times \theta_{i,\gbest}.
\end{equation}

We present the error in various parameters in the fifth column of Table~(\ref{tab:results}).
\subsection{Two dimensional contour plots}
\begin{figure*}
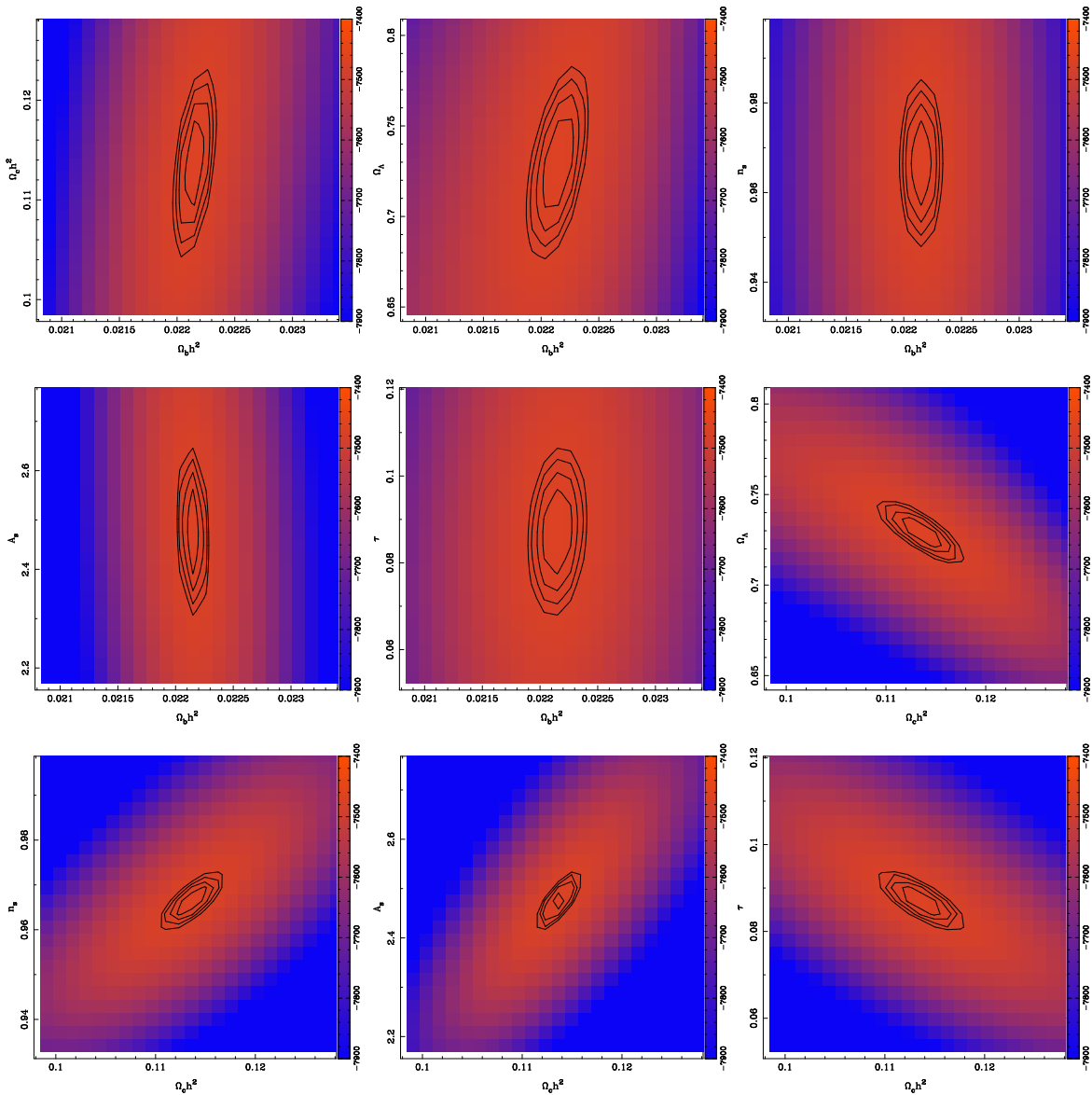

\begin{center}
\begin{tabular}{ccc}
\epsfig{width=5cm,angle=-90,file=plot_20.ps} &
\epsfig{width=5cm,angle=-90,file=plot_21.ps} &
\epsfig{width=5cm,angle=-90,file=plot_22.ps}  \\
\epsfig{width=5cm,angle=-90,file=plot_23.ps} &
\epsfig{width=5cm,angle=-90,file=plot_24.ps} &
\epsfig{width=5cm,angle=-90,file=plot_25.ps}  \\
\epsfig{width=5cm,angle=-90,file=plot_26.ps} &
\epsfig{width=5cm,angle=-90,file=plot_27.ps} &
\epsfig{width=5cm,angle=-90,file=plot_28.ps}  \\
\end{tabular}
\caption{Panels in this figure show the two dimensional contour plots for various pairs of
the six cosmological parameters $\Omega_bh^2,\Omega_ch^2,\Omega_{\Lambda},n_s,A_S$ and $\tau$. 
We have considered a grid of size $24\times 24$ for our exercise considered the $3\sigma$ region,
(where $\sigma$ is the error computed from the fitting) around the best fit point. In the case 
when we do not have any idea about the $\sigma$ we can give any other trial value also. The contours 
in the figure (from inside) are for $\Delta \chi^2=2,4,6,8 $ and 10. Note that the contour plots not
only give an idea about the area around the best fit region, they also clearly demonstrate 
correlation between various cosmological parameter.
}
\label{fig:cont1}
\end{center}
\end{figure*}
\begin{figure*}
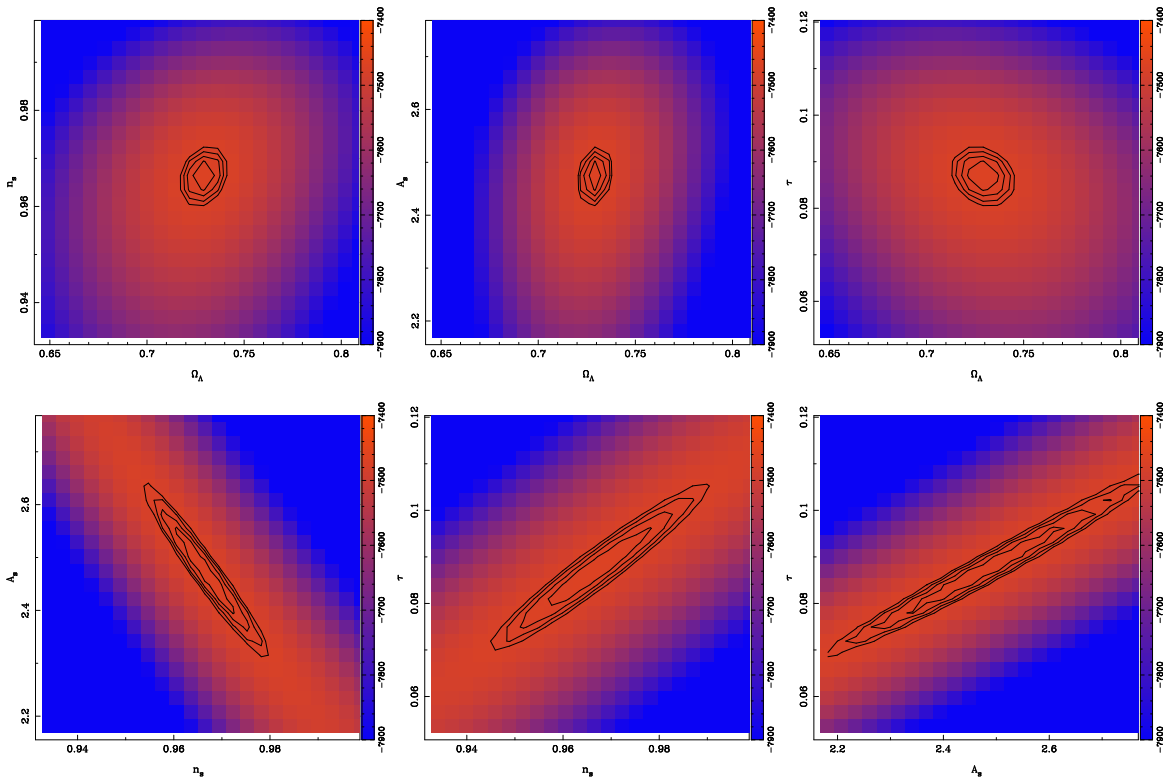

\begin{center}
\begin{tabular}{ccc}
\epsfig{width=5cm,angle=-90,file=plot_29.ps} &
\epsfig{width=5cm,angle=-90,file=plot_30.ps} &
\epsfig{width=5cm,angle=-90,file=plot_31.ps}  \\
\epsfig{width=5cm,angle=-90,file=plot_32.ps} &
\epsfig{width=5cm,angle=-90,file=plot_33.ps} &
\epsfig{width=5cm,angle=-90,file=plot_34.ps}  \\
\end{tabular}
\caption{Same as in Figure~\ref{fig:cont1} but for different combination of cosmological parameters.}
\label{fig:cont2}
\end{center}
\end{figure*}

Fitting as we have done may not give a very good estimate of the errors on parameters. 
The exact way to figure out how the likelihood surface behaves around the best fit location,
which is given by \gbest~ in our case is to compute  the likelihood function 
on a grid around the best fit point. Numerical computation over a multi-dimensional grid 
is quite expensive and even for a moderate size grid of 24 we have to do $24^6$ computations
for a six parameter cosmological model. In place of considering a multi-dimensional grid we consider,
$N_d(N_d-1)/2$ two dimensional grid as is done by some other authors  \cite{2003ApJ...596..725C}.
We fix the values of the four parameters out of six to their best fit values i.e., \gbest~ and
draw the two dimensional grid over the other  two parameters.

In Figure~\ref{fig:cont1} and \ref{fig:cont2} we show two dimensional contour plots for 
different pairs of cosmological parameters.  The contours from the innermost are for
$\Delta \chi^2=2,6,8$ and $10$.  Note that the range for  grids  is selected from the 
rough estimates of errors which we get from fitting. The range for $\Omega_ch^2$ and 
$\Omega_{\Lambda}$ is taken $2\sigma$ and for others it is taken $3\sigma$ where 
$\sigma$ is the error obtained from the fitting

\subsection{Primordial Power Spectrum (PPS) with power in bins}

\begin{figure}
\begin{center}
\epsfig{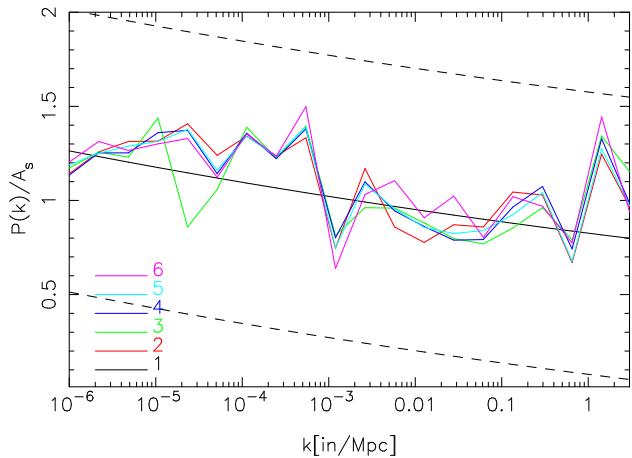}
\caption{
This figure shows how the binned primordial power spectrum 
(20 logarithmic  bins over the $k$ range) changes as PSO 
progresses (line 1 is for the initial PPS and 6 is for the final PPS). 
The lower and upper values of the power in bins are represented by the dashed line.
Starting with a power law PPS we found that a power spectrum which has low
power in some bins and high in others fits better than a 
power law model.}
\label{fig:pps}
\end{center}
\end{figure}

\begin{figure}
\begin{center}
\epsfig{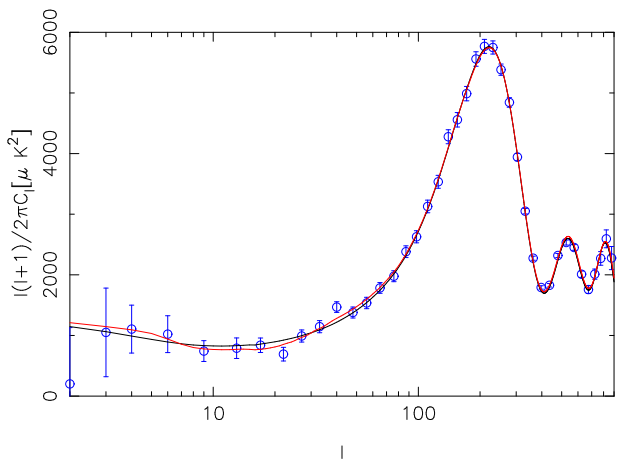}
\caption{The red, black and blue lines in the above figure represent the best fit angular power spectrum
recovered from PSO, standard LCDM power spectrum  and the binned power spectrum of WMAP seven 
year data respectively.  Note that at low $l$ the angular power spectrum with binned PPS fits better 
as compared to the standard power law PPS to the observed data (the improvement in $\Delta\chi^2$ is around 7).
}
\label{fig:cls_pps}
\end{center}
\end{figure}

In order to demonstrate an interesting example in which PSO can be useful, we consider
a model in which the primordial power spectrum has ``binned" power in place of being a 
power law. We consider the power in bins as  free parameters and that make our model 
higher dimensional. Apart from having 20 free parameters, which gives power in logarithmic
bins we consider rest of the four parameters $\Omega_bh^2, \Omega_ch^2, \Omega_{\Lambda}$ and 
$\tau$ also free. As is expected,  a model with more parameters better fits the observational 
data, which in the present case is WMAP seven year data i.e., for a primordial power 
spectrum with binned power $\chi^2$ is lower by 7 as compared to the standard power law mode.
The best fit primordial power spectrum and angular power spectrum are shown in Figure \ref{fig:pps}
and Figure \ref{fig:cls_pps} respectively. Note that the PPS with binned power gives better fit,
particularly at low l. We consider here the particular form of PPS not to motivate any
particular theoretical model but just to demonstrate the method we use.

\subsection{Computational performance}
Computing the fitness function i.e., $-2 \log~{\mathcal L}$,  which is the most expensive 
part in the parameter estimation procedure, involves two steps. 
In the first step the $C_l$s are computed for a set of cosmological parameters for 
which we use the publicly available code CAMB~\cite{2000ApJ...538..473L} which employs 
OpenMP pragmas for doing computationally intensive steps in parallel on multi-processor shared 
memory systems. 
In the second step the likelihood function is computed from  $C_l$s and the observational
data i.e., WMAP seven year data, for which we use the likelihood code provided by the WMAP team. 
Since our PSO code shares two main modules ($C_l$ and likelihood computation) with the publicly available 
code COSMOMC,  the difference in the performance is expected only due to the number of times the fitness
function is computed. Computationally a typical PSO run gives very good convergence with 30 PSO 
particles with 160 steps i.e., 4800 computations. We ran a typical COSMOMC run with  24 chains
and found that (from reading .log file for every chain) that CAMB was called 691200 times,
which is more than  50 times the number of CAMB  calls made in our case (including the calls
for two dimensional grids used for contours). 

Here it is also important to mention that parallelizing our code is very straightforward. 
We use OpenMP to compute $C_l$ for a point in the six dimensional parameter space and 
use MPI to distribute particles among different MPI nodes.  At every step, particles are 
distributed among MPI nodes and after they return the value of the fitness function, the master
node computes  ~\pbest~ and ~\gbest~ and updates the positions and velocities of particles.
We have tested our code on a Linux cluster using 15 MPI nodes,  where each node has 
2 AMD quad core Opetron 2.6 GHz processors.  Within a node we use OpenMP for computing 
the angular power spectrum using as many number of threads as the number of cores are present. 
We have also ported our code on a Cray CX1 system with six nodes, where each node has 
2 Intel Xeon 2.67 Ghz  6 core processors.  In a case when the number of nodes can 
divide the number of PSO particles,  there is no difficulty with load balance. 
Since at every PSO step a very small amount of data has to be communicated  among MPI nodes,
we have found that MPI collective communications calls like ``broadcast" and ``gather"
are much more efficient than regular ``send" and ``receive" calls.
In the present run, the result of which are reported here, it took roughly two and half hours for 
the standard PSO run to finish on a  Linux cluster with 10 nodes with each  node having  
2 AMD quad core Opetron 2.6 GHz processors.  The convergence was found just after 159 steps 
with 30 PSO particles.

It is not very straightforward to compare the performance of our PSO code and 
that of the commonly used code COSMOMC mainly because:
\begin{enumerate}
\item
The convergence criteria in PSO is slightly different than that in  COSMOMC. 
\item
The angular power spectrum $C_l$ computation in 
COSMOMC is optimized by selecting only a subset of particles which change their value,
i.e., fast-slow parameters. There  is no such operation in PSO.
\item
It is a common practice to ``thin" chains  in MCMC which means that not all sampled 
points are used for the final result that is not the case in PSO.
\item
Inputs for PSO and COSMOMC are different, in the sense COSMOMC needs a 
a guess covariance matrix, widths of the final one dimensional probability 
distribution and a starting point, apart from the search range. Which is not 
the case in PSO in which we only need to specify a reasonable search range. 
\end{enumerate}

In \cite{2010LNP...800..147V} four chains with each having 30,000 points are 
needed for convergence, but in our typical run with six parameter model we never need 
more than 8-9,000 computations. Here it is important to note that the convergence 
in PSO also depend on value of design parameters.

\section{Discussion and conclusion}
\label{ss:pso_summary}
In the present work we have demonstrated the application of Particles Swarm Optimization or PSO 
for cosmological parameter estimation from CMB data, which we believe has not been 
done earlier in any other study. Being a different method, PSO can be used as 
an alternative technique  for parameter estimation, particularly when one is mainly 
interested in the location of the best fit point. 
The main focus of our present work was to demonstrate the method, in 
place of producing quantitative results and comparing those with other methods,  
that  we leave for our future work. We have not only shown how to compute the 
values of the best-fit parameters, but  have  also proposed a method to quantify the error bars 
on the estimated values. 

Based on a very simple algorithm, PSO has many interesting features some  
are as follows:
\begin{enumerate}
\item
PSO has very few design parameters , the values of which can be easily fixed. 
\item
By tuning the values of the design parameters, PSO can be made  more efficient for  global 
or a local search although it is more useful for a global search.
\item
In PSO no approximation or extrapolation is made at any step (like in artificial neural network) 
and the optimization function is computed exactly at every point. 
\item
As is claimed in other studies also  PSO is very efficient in searching 
for the global maximum when dimensionality of the search space is very high or
there are a large number of local maxima present.  Adding extra search dimensions 
in PSO is quite straightforward. 
\item
PSO can be used for accelerated search of the global maximum since it always has 
some idea about the \gbest~ from the very beginning. 
\item
In PSO we need to give only the search range as an input and no other 
information (as is needed in MCMC) about parameters, like covariance matrix, width of the final
1-d probability distribution or starting point is needed. 
This can  be very useful for models about the parameters of which we do not
have much prior information.
\item
Since this method is also based on Bayesian formalism so extra data sets can 
be easily incorporated in this method also.
\end{enumerate}

As a result of very high quality CMB data which has already been provided by
the WMAP satellite~\cite{wmap} and will be provided by the Planck satellite~\cite{planck},
it has become an interesting exercise to consider much more complex models (with very high 
dimensionality than just six to eleven dimensional models). 
The main motivation behind considering such models has been to fit the 
``outliers" of the WMAP data (all years).  In one of such exercise, in place of 
considering the primordial power spectrum just a power law, power in various bins 
can be left open for the fitting, which makes  the dimensionality of the search space very large
and and we have shown that  PSO can be quite useful for such problems. 

In the present work we have demonstrated that particle swarm optimization can also be 
used for cosmological parameter estimation from  CMB data sets.
Apart from discussing the technique in detail,  we have also presented our results 
for the standard six parameters cosmological model. Along with giving the best 
fit parameters for the WMAP seven year data, we have also given some rough estimates
of the errors, and have shown two dimensional contour plots in order to make the 
treatment complete.
We have also shown an application of our method for a higher dimensional cosmological model
in which the primordial power spectrum has power in logarithmic bins as free parameters.
The main aim of the present work was to demonstrate a new method and present the results
qualitatively. In future we plan to present a detailed quantitative analysis.

\bibliographystyle{h-physrev3}
\bibliography{cmbr}

\begin{thebibliography}{10}

\bibitem{1996PhRvD..54.1332J}
G.~{Jungman}, M.~{Kamionkowski}, A.~{Kosowsky}, and D.~N. {Spergel},
\newblock \prd {\bf 54}, 1332 (1996), arXiv:astro-ph/9512139.

\bibitem{1996PhRvL..76.1007J}
G.~{Jungman}, M.~{Kamionkowski}, A.~{Kosowsky}, and D.~N. {Spergel},
\newblock Physical Review Letters {\bf 76}, 1007 (1996),
  arXiv:astro-ph/9507080.

\bibitem{2002PhRvD..66j3511L}
A.~{Lewis} and S.~{Bridle},
\newblock \prd {\bf 66}, 103511 (2002), arXiv:astro-ph/0205436.

\bibitem{2002PhRvD..66f3007K}
A.~{Kosowsky}, M.~{Milosavljevic}, and R.~{Jimenez},
\newblock \prd {\bf 66}, 063007 (2002), arXiv:astro-ph/0206014.

\bibitem{2003ApJS..148..195V}
L.~{Verde} {\em et~al.},
\newblock \apjs {\bf 148}, 195 (2003), arXiv:astro-ph/0302218.

\bibitem{2010LNP...800..147V}
L.~{Verde},
\newblock {Statistical Methods in Cosmology},
\newblock in {\em Lecture Notes in Physics, Berlin Springer Verlag}, edited by
  {G.~Wolschin}, volume 800 of {\em Lecture Notes in Physics, Berlin Springer
  Verlag}, pp. 147--177, 2010, 0911.3105.

\bibitem{2003moco.book.....D}
S.~{Dodelson},
\newblock {\em {Modern cosmology}} ({Academic Press}, {San Diego, U.S.A.},
  2003).

\bibitem{2011ApJS..192...14J}
N.~{Jarosik} {\em et~al.},
\newblock \apjs {\bf 192}, 14 (2011), 1001.4744.

\bibitem{2011ApJS..192...16L}
D.~{Larson} {\em et~al.},
\newblock \apjs {\bf 192}, 16 (2011), 1001.4635.

\bibitem{2001CQGra..18.2677C}
N.~{Christensen}, R.~{Meyer}, L.~{Knox}, and B.~{Luey},
\newblock Classical and Quantum Gravity {\bf 18}, 2677 (2001),
  arXiv:astro-ph/0103134.

\bibitem{2007MNRAS.376L..11A}
T.~{Auld}, M.~{Bridges}, M.~P. {Hobson}, and S.~F. {Gull},
\newblock \mnras {\bf 376}, L11 (2007), arXiv:astro-ph/0608174.

\bibitem{KennedEberhart1995}
J.~{Kennedy} and R.~C. {Eberhart},
\newblock IEEE International Conference on Neural Networks {\bf 4}, 1992
  (1995).

\bibitem{KennedEberhart2001}
J.~{Kennedy} and R.~C. {Eberhart},
\newblock {\em {Swarm Intelligence}} ({Morgan Kufmann}, 2001).

\bibitem{Engelbrech2002}
A.~P. {Engelbrecht},
\newblock {\em {Computational Intelligence, An Introduction}} ({John Wiley \&
  Son}, 2002).

\bibitem{lambda}
http://lambda.gsfc.nasa.gov/.

\bibitem{2010PhRvD..81f3002W}
Y.~{Wang} and S.~D. {Mohanty},
\newblock \prd {\bf 81}, 063002 (2010), 1001.0923.

\bibitem{2010JCAP...10..008H}
D.~K. {Hazra}, M.~{Aich}, R.~K. {Jain}, L.~{Sriramkumar}, and T.~{Souradeep},
\newblock \jcap {\bf 10}, 8 (2010), 1005.2175.

\bibitem{2011arXiv1106.2798A}
M.~{Aich}, D.~K. {Hazra}, L.~{Sriramkumar}, and T.~{Souradeep},
\newblock ArXiv e-prints  (2011), 1106.2798.

\bibitem{2011arXiv1109.5264M}
P.~D. {Meerburg}, R.~{Wijers}, and J.~P. {van der Schaar},
\newblock ArXiv e-prints  (2011), 1109.5264.

\bibitem{Lazinic2009}
A.~{Lazinica},
\newblock {\em {Particle Swarm Optimization}} ({In-Tech.}, 2009).

\bibitem{Boeringer2004}
D.~W. {Boeringer} and D.~H. {Werne},
\newblock IEEE Transc. on Antennas and propagation {\bf 52}, 771 (2004).

\bibitem{Robinson2004}
J.~{Robinson} and Y.~{Rahmat-Samii},
\newblock IEEE Trans. on Ant. and Prop. {\bf 727}, 397 (2004).

\bibitem{2005MNRAS.359..251S}
C.~{Skokos}, K.~E. {Parsopoulos}, P.~A. {Patsis}, and M.~N. {Vrahatis},
\newblock \mnras {\bf 359}, 251 (2005), arXiv:astro-ph/0502164.

\bibitem{2011ApJ...727...80R}
A.~{Rogers} and J.~D. {Fiege},
\newblock \apj {\bf 727}, 80 (2011), 1101.5803.

\bibitem{pso2006}
www.particleswarm.info/Standard\_PSO\_2006.c/.

\bibitem{1992GelmanRubin}
A.~{Gelman} and R.~D.B,
\newblock Statistical Science {\bf 7}, 457 (1992).

\bibitem{1998BrooksGelman}
S.~P. {Brooks} and G.~A.,
\newblock Journal of Computational and Graphical Statistics {\bf 7}, 434
  (1998).

\bibitem{1984ApJ...285L..45B}
J.~R. {Bond} and G.~{Efstathiou},
\newblock \apjl {\bf 285}, L45 (1984).

\bibitem{2004astro.ph..3344C}
A.~{Challinor},
\newblock arXiv:astro-ph/0403344  (2004), arXiv:astro-ph/0403344.

\bibitem{1996LNP...470..207H}
W.~{Hu},
\newblock {Concepts in CMB Anisotropy Formation},
\newblock in {\em The Universe at High-z, Large-Scale Structure and the Cosmic
  Microwave Background}, edited by {E.~Martinez-Gonzalez \& J.~L.~Sanz}, volume
  470 of {\em Lecture Notes in Physics, Berlin Springer Verlag}, pp. 207--+,
  1996, arXiv:astro-ph/9511130.

\bibitem{1987MNRAS.226..655B}
J.~R. {Bond} and G.~{Efstathiou},
\newblock \mnras {\bf 226}, 655 (1987).

\bibitem{1997MNRAS.291L..33B}
J.~R. {Bond}, G.~{Efstathiou}, and M.~{Tegmark},
\newblock \mnras {\bf 291}, L33 (1997), arXiv:astro-ph/9702100.

\bibitem{1998PhRvD..57.2117B}
J.~R. {Bond}, A.~H. {Jaffe}, and L.~{Knox},
\newblock \prd {\bf 57}, 2117 (1998), arXiv:astro-ph/9708203.

\bibitem{2003ApJ...596..725C}
M.~{Chu}, M.~{Kaplinghat}, and L.~{Knox},
\newblock \apj {\bf 596}, 725 (2003), arXiv:astro-ph/0212466.

\bibitem{camb}
http://camb.info/.

\bibitem{2000ApJ...538..473L}
A.~{Lewis}, A.~{Challinor}, and A.~{Lasenby},
\newblock \apj {\bf 538}, 473 (2000), arXiv:astro-ph/9911177.

\bibitem{1996ApJ...469..437S}
U.~{Seljak} and M.~{Zaldarriaga},
\newblock \apj {\bf 469}, 437 (1996), arXiv:astro-ph/9603033.

\bibitem{2008PhRvD..77j3013H}
S.~{Hamimeche} and A.~{Lewis},
\newblock \prd {\bf 77}, 103013 (2008), 0801.0554.

\bibitem{2009PhRvD..79h3012H}
S.~{Hamimeche} and A.~{Lewis},
\newblock \prd {\bf 79}, 083012 (2009), 0902.0674.

\bibitem{wmap}
http://map.gsfc.nasa.gov/.

\bibitem{gsl}
http://www.gnu.org/software/gsl/.

\bibitem{planck}
http://www.nasa.gov/mission\_pages/planck/.

\end{thebibliography}

\end{document}